\documentclass[%
  twocolumn,%
  groupedaddress,%
  letterpaper,%
  amsfonts,%
]{revtex4}

\usepackage{amsmath,amssymb,bm}
\usepackage[pdftex]{graphicx}
\usepackage{subfigure}
\usepackage{hyperref}
\hypersetup{%
  breaklinks = {true},
  citecolor = {blue},
  colorlinks = {true},
  linkcolor = {red},
  pdfauthor = {\textcopyright\ Vitor M. Pereira },
  pdfcreator = {\LaTeX\ and \flqq hyperref\frqq},
  pdffitwindow = {true},
  pdfmenubar = {true},
  pdfpagelayout = {SinglePage},
  pdfstartview = {Fit},
  pdftoolbar = {true},
  plainpages = {false},
}

\newcommand{\vF}{{\ensuremath{v_\text{F\,}}}}

\newcommand{\Acal}{\ensuremath{\mathcal{A}}}
\newcommand{\Acalb}{\ensuremath{\bm{\mathcal{A}}}}

\newcommand{\bn}{\ensuremath{\bm{n}}}
\newcommand{\bnabla}{{\ensuremath{\bm{\nabla}}}}
\newcommand{\bk}{{\ensuremath{\bm{k}}}}

\newcommand{\bd}{{\ensuremath{\bm{d}}}}
\newcommand{\bp}{{\ensuremath{\bm{p}}}}
\newcommand{\bq}{{\ensuremath{\bm{q}}}}
\newcommand{\br}{{\ensuremath{\bm{r}}}}
\newcommand{\bR}{{\ensuremath{\bm{R}}}}

\newcommand{\bu}{{\ensuremath{\bm{u}}}}
\newcommand{\bz}{{\ensuremath{\bm{z}}}}
\newcommand{\ce}{{\ensuremath{\varepsilon}}}

\newcommand{\Fref}[1]{Fig.~\ref{#1}}
\newcommand{\Eqref}[1]{Eq.~\eqref{#1}}
\newcommand{\Ref}{Ref.~}

\begin{document}


\title{Geometry, mechanics and electronics of singular structures and
wrinkles in graphene}

\author{Vitor M. Pereira}
\thanks{Corresponding author. Author email: \texttt{vpereira@bu.edu}.}
\affiliation{Department of Physics, Boston University, 590
Commonwealth Avenue, Boston, MA 02215, USA}

\author{A.~H. Castro Neto}
\affiliation{Department of Physics, Boston University, 590
Commonwealth Avenue, Boston, MA 02215, USA}

\author{H. Y. Liang}
\affiliation{School of Engineering and Applied Sciences, Harvard University,
29 Oxford Street, Cambridge, Massachusetts 02138, USA}

\author{L. Mahadevan}
\thanks{Corresponding author. Author email: \texttt{lm@seas.harvard.edu}.}
\affiliation{School of Engineering and Applied Sciences, Harvard University,
29 Oxford Street, Cambridge, Massachusetts 02138, USA}

\date{\today}


\begin{abstract}
As the thinnest atomic membrane, graphene presents an opportunity to combine
geometry, elasticity and electronics at the limits of their validity. The
availability of reliable atomistic potentials for graphene allows
unprecedented precise simulations of the mechanical response of atomic
membranes. Here we
describe the transport and electronic structure in the neighbourhood of conical
singularities, the elementary excitations of the ubiquitous wrinkled and
crumpled graphene that occur in non-epitaxial suspended samples where
shear stresses are unavoidable. We use a combination of atomistic mechanical
simulations, analytical geometry and transport calculations in curved graphene,
and exact diagonalization of the electronic spectrum to calculate the effects of
geometry on electronic structure, transport and mobility in suspended samples.
We also point out how the geometry-generated pseudo-magnetic/electric fields
might disrupt Landau quantization under a magnetic field. 
\end{abstract}

\maketitle


%
%
%
Graphene wrinkles easily and often \cite{Meyer:2007}. This effect is most
clearly observed in samples obtained from exfoliation of graphite, and
subsequent deposition onto a substrate \cite{Lau:2009}, or in chemically derived
oxides \cite{Stankovich:2007}. Since graphene is an atomically thin membrane, 
it is impossible to lay a shear-free sheet of it onto a flat surface, as it
sticks almost immediately to a substrate --- such as the edges of a trench via
van der Waals interaction ---, and the substrate is itself rarely, if ever, flat
\cite{Stolyarova:2007,Ishigami:2008,Geringer:2009}, so that perfect shear-free
conformations are not possible. In addition, recently advanced techniques to
grow graphene on metallic surfaces clearly show widespread wrinkling arising
from thermal expansion mismatch between graphene and the host substrate
\cite{Bae:2009,Sutter:2009}.
These boundary deformations acting on graphene
lead to wrinkling because of the nearly negligible threshold for 1D and 2D
buckling instabilities in thin plates; the bending rigidity scales with the cube
of the thickness so that  a thin membrane cannot support even arbitrarily small
shear or compression without wrinkling on scales large compared to its thickness
\cite{Landau-Elasticity}.

However, for all its flexural limpness, graphene exhibits the largest in-plane
Young's modulus \cite{Lee:2008} and, though easy to bend, is extremely hard to
stretch. This geometry-induced separation of the energy scales for thin
membranes implies that they try to respond to shear by bending isometrically
almost everywhere \cite{Cerda:2004}. However, except in very limited cases
corresponding to developable deformations, bending alone cannot accommodate the
state of stress or the boundary conditions imposed by the geometry. This
conflict is resolved naturally by local membrane stretching by an amount
sufficient to just accommodate the imposed geometric and physical constraints,
so that regions of in-plane strain are restricted to vanishingly small areas
distributed throughout the system. A simple example is seen in a thin sheet of 
paper which is very resistant to stretching (it actually tears before we can
stretch it), but bends easily; when a piece of crumpled paper is straightened
out, we see flat areas connected by a network of ridges that meet at sharp
vertices: the highly localized scars where the sheet is plastically deformed.
The peaked structures constitute the basic element of the entire ridge network,
and serve to focus large strains and energy densities. They are ubiquitous in
thin films that are strongly deformed in such instances as drapes
\cite{Cerda:2004}, skin winkles, etc., and are termed developable cones or
conical singularities (CSs); their outer form has been geometrically and
mechanically characterized in terms of a theory for the inextensional
deformations of thin sheets \Ref\cite{Cerda:1998, Cerda:2005}. In particular,
there is a simple universal analytic expression for their geometry as a function
of the boundary and/or stress conditions on the sheet far from the nearly
singular tip where the effects of stretching are concentrated.

Exfoliated graphene and suspended samples \cite{Lau:2009} naturally exhibit CSs
as shown in \Fref{fig:Fig-1}(a), which are of particular interest in the quest
for ultimate electronic mobility \cite{Bolotin:2008,Li:2009} and non-trivial
interaction effects \cite{Du:2009,Bolotin:2009}. Here we build on our knowledge
of the structure and mechanics of CSs to study the influence of these ubiquitous
objects on electronic transport in graphene. Unlike in most solid-state
materials, flexural and planar deformations couple to electrons in graphene in a
peculiar way. The honeycomb lattice implies that the effective electronic
excitations of the system are two dimensional massless Dirac fermions
\cite{RMP:2009}. The geometry of and strain in the lattice then couples to these
excitations through both effective gauge fields, and local scattering potentials
that follow the local curvature and thus affect the electronic structure
\cite{Pereira:2009b,Ribeiro:2009}, opening the prospect for strain-engineered
electronic devices \cite{Pereira:2009a,Guinea:2009}. CSs are also present in
buckled nanotubes, where they have been shown to significantly alter transport
characteristics \cite{NanotubesBending}.

%
%
\section{Conical Singularities}
Graphene does behave like a thin plate under stress, even at the atomistic
level; when sheared biaxially, and afterwards allowed to relax via Molecular
Dynamics (MD) \cite{SuppInfo}, \Fref{fig:Fig-1}(c) shows the relaxed
configuration, which exhibits the classical Miura-ori like ridge pattern of 2D
buckling \cite{Mahadevan:2005}, with the CSs arising at the intersection of the
ridges \Fref{fig:Fig-1}(e,f). 

In cylindrical coordinates, the displacement field associated with the CSs, a
generalized cone, reads $\bu(\rho,\theta)  = \br-\br_0 = u_\rho(\rho,\theta)
\bu_ \rho + u_\theta(\rho, \theta) \bu_\theta + \zeta(\rho,\theta) \bz$, where
$\zeta(\rho,\theta) = \rho \psi(\theta)$. The solution for
$u_\theta(\rho,\theta), u_\rho(\rho,\theta)$, and $\psi(\theta)$ is obtained by
solving the equations of equilibrium for the finite bending of a plate  with the
constraint of inextensibility, i.e. that there is no in-plane strain
($\gamma_{ij}=0$). The vertical displacement is then given by (Refs.
\cite{Cerda:1998,SuppInfo}) 
\begin{equation}
  \zeta = \rho\psi(\theta), \
  \psi(\theta) = \ce \Theta(|\theta| - \theta_1) 
               + \ce \Psi(\theta) \Theta(\theta_1 - |\theta|)
  \label{eq:psi} 
  \,;
\end{equation}
where $\ce$ characterizes the angle of the enveloping cone, and both
$\theta_1\approx 70^\text{o}$ and $\Psi(\theta)$ are universal. This
independence of the shape on any material parameters and scale, together with
the Cauchy-Born hypothesis, allows us to describe conical singularities and
wrinkling in graphene by applying the deformation field $\bu(\rho,\theta)$ to
all atoms in the lattice. The resulting shape of the lattice is the one shown in
\Fref{fig:Fig-1}(e), with the main effects arising from curvature. Since
$\bu(\rho,\theta)$ is constructed so that there is no in-plane strain; however
some localized stretching strain is concentrated in the neighbourhood of the
apex
which  will relax naturally in a MD simulation as a consequence of the large
but finite stretching rigidity of graphene so that, even after relaxation, all
inter-atomic distances are strongly peaked about the natural lattice spacing
$a=a_0=1.42\text{\AA}$, as can be seen in \Fref{fig:Fig-1}(d), with a spread of 
$~2\%$ for the values of $\ce$ of interest here. This is just a reflection of
the relative inextensibility of the in-plane $\sigma$ bonds, which leads to a
blunting of the apex but is of little significance elsewhere. Since the relaxed
structure shows strain $>1\%$ for a dozen of  atoms only, and very near the
apex, we shall neglect it altogether.

%
%
\section{Effective Model}
To understand how  the electronic properties respond to this deformation, we
note that the relevant physics occurs in the $p_z$-derived $\pi$ bands of
graphene; curvature causes re-hybridization of these orbitals
\cite{EunAhKim:2008}, hindering or favouring wavefunction overlap, and thus
perturbs the electronic kinetic energy. This affects both the $\pi$ band
sub-system and hybridizes the $p_z$ and the $sp^2$ sub-bands, which are
otherwise orthogonal. As a first step we shall neglect this latter effect, which
mostly shifts the chemical potential, and focus only on the $\pi$ bands.

Within the tight-binding approximation, the bandstructure is then determined by
the effective Hamiltonian 
\begin{equation}
  H = \sum_{\langle i,j \rangle} t_{ij} c^\dagger_ic_j +
      \sum_{\langle\langle i,j \rangle\rangle} 
        t'_{ij} c^\dagger_ic_j + \text{H. c.}
  \label{eq:H-tb}
  \,,
\end{equation}
where the two contributions come from first and second neighbours, and
$t^{(\prime)}_{ij}=V_{pp\pi}$ is the two centre Slater-Koster overlap integral
\cite{Slater:1954}, which has to be calculated now for all pairs of neighbours,
taking into consideration the full geometry of the deformed lattice. To do this,
we introduce the unit normal at every point of the surface, $\bn(\rho,\theta)$,
so that for two atoms separated by an arbitrary distance $\bd=\bR_i- \bR_j$,
straightforward rotation of the $p_z$ orbitals and Slater-Koster tables tell us
that the overlap integral is \cite{Isacsson:2008}
\begin{equation}
  t_{ij} = V_{pp\pi} \, \bn_i\cdot\bn_j + 
           \bigl( V_{pp\sigma} - V_{pp\pi} \bigr)
            (\bn_i\cdot\hat{\bd})(\bn_j\cdot\hat{\bd})
  \label{eq:Hopping}
  \,.
\end{equation}
Since the surface is completely parametrized by the normal displacement field,
we may use the geometry of the developable cone to obtain the normals, distances
and the hopping $t_{ij}$ among any two atoms, noting that the underlying metric
remains Euclidean. To make progress analytically we assume, with Harrison
\cite{Harrison:1999}, that $d^2V_{ppx}(d) = d_0^2V_{ppx}(d_0)$, so that on
solving the Gauss equations we obtain $t_{ij}(\bd) = t^0_{ij}(\bd_0) + \delta
t_{ij}$, with:
\begin{equation}
  \delta t_{ij} \approx - V_{pp\pi}^0 \frac{1}{2}
    \bigl|(\bd_0\cdot\bnabla)\bnabla\zeta\bigr|^2
    + \frac{V_1}{d_0^2} \bigl[ (\bd_0\cdot\bnabla)^2 \zeta \bigr]^2
  ,
  \label{eq:delta-t}
\end{equation}
$V_1 =V_{pp\pi}^0/3 - V_{pp\sigma}^0/4$. In the low energy approximation, we may
then  describe \eqref{eq:H-tb} by the effective Dirac Hamiltonian:
$
  H \approx v_F \bm{\sigma}.
        \bigl[ \bp-\tfrac{1}{\vF}\bm{\mathcal{A}(\br)} \bigr]
        + \bigl[3 t'_0 + \Phi(\br)\bigr] \, \sigma^0
  \,
$
in each valley of the Brillouin zone. Then the effective gauge field
$\Acalb(\br)$ and the local potential $\Phi(\br)$ depend, respectively, on the
perturbations of the nearest neighbour, and next-nearest neighbour hopping
\Ref\cite{RMP:2009} via
\begin{equation}
  \Acal_x\!-\! i \Acal_y = \sum_{\bn} \delta t_{\bn}(\br) \, 
              e^{i\bk\cdot\bn}
  ,\
  \Phi = \sum_\Delta \delta t'_\Delta(\br) \,
              e^{i\bk\cdot\bm{\Delta}} 
  .
  \label{eq:Fields}
\end{equation}
Substituting \eqref{eq:delta-t} in \eqref{eq:Fields}, one obtains
\begin{equation}
  \Phi(\br) = \alpha\, \text{Tr}^2 \bigl[
              \partial^i\partial_j\zeta \bigr]
            - \beta\, \text{det}
              \bigl[ \partial^i\partial_j\zeta \bigr]
  \label{eq:Phi}
  ,
\end{equation}
with $\alpha=9a_0^2V^0_{pp\pi}/8 + 27a_0^2V_{pp\sigma}/32 \approx
1.5\text{eV\AA}^2$, and $\beta=3a_0^2V^0_{pp\pi}+9a_0^2V^0_{pp\sigma}/8 \approx
3\text{eV\AA}^2$ \cite{Endnote-2}. We recall that 
$\partial^i\partial_j\zeta\approx K^i_{\phantom{i}j}$  is the curvature tensor
of our conical surface and, since  $H=\text{Tr}K^i_{\phantom{i}j}/2$ and
$\text{det}K^i_{\phantom{i}j}$ are the local mean and Gaussian curvatures, it
follows that $\Phi$ is entirely determined by the cone geometry. Moreover, since
CSs are developable surfaces, the Gaussian curvature vanishes everywhere, so
that $\Phi(\br)=\alpha a_0^2 (\nabla^2\zeta)^2$. The gauge field $\Acalb$ is
also given in terms of products of $\partial^i\partial_j\zeta$, but we shall not
write it explicitly since this potential couples to the electric current, and 
therefore does not contribute in leading order for scattering and transport when
time-reversal symmetry is preserved. However $\Phi$ leads to an electrostatic
potential that is felt by the Dirac electrons and thus contributes directly to
the resistivity.

%
%
\section{Transport}
We now consider the contribution of the CSs for the momentum relaxation time in
the Boltzmann formalism. In the Born approximation, the scattering rate is given
by $S(\bk,\bk') = 2\pi/\hbar |V_{\bk,\bk'}|^2 \delta (E_\bk-E_{\bk'})$, with 
$V_{\bk,\bk'} = \Phi_{\bk-\bk'}[1+\exp(i\phi_\bk-i\phi_{\bk'})]/2$, and
$\Phi_\bq$ is the Fourier transform of the local potential \eqref{eq:Phi}:
$\Phi(\br) = \alpha a_0^2 [\psi(\theta)+\psi''(\theta)] / r^2$, which is of
course directly related to the cone geometry. This potential is unusual for two
reasons: it is anisotropic on account of \eqref{eq:psi} and decays in space as $
\propto 1/r^2$,  so that it is beyond the supercritical threshold for Dirac
fermions in 2D \cite{Novikov:2007}. Were it not for the natural lattice 
regularization at $r\sim0$, such potential would lead to an unbound spectrum of
discrete states. This effect is also blunted the mechanical, stretch-induced
relaxation observed near the apex in \Fref{fig:Fig-1}(e,f). The
result is a short range potential with a finite number of bound states (unlike
the Coulomb case where the long range $1/r$ tail begets an infinite spectrum of
resonances, even after regularization), so that CSs therefore scatter as short
range, anisotropic potentials.

\begin{figure*}[h]
  \centering
  \includegraphics[width=0.8\textwidth]{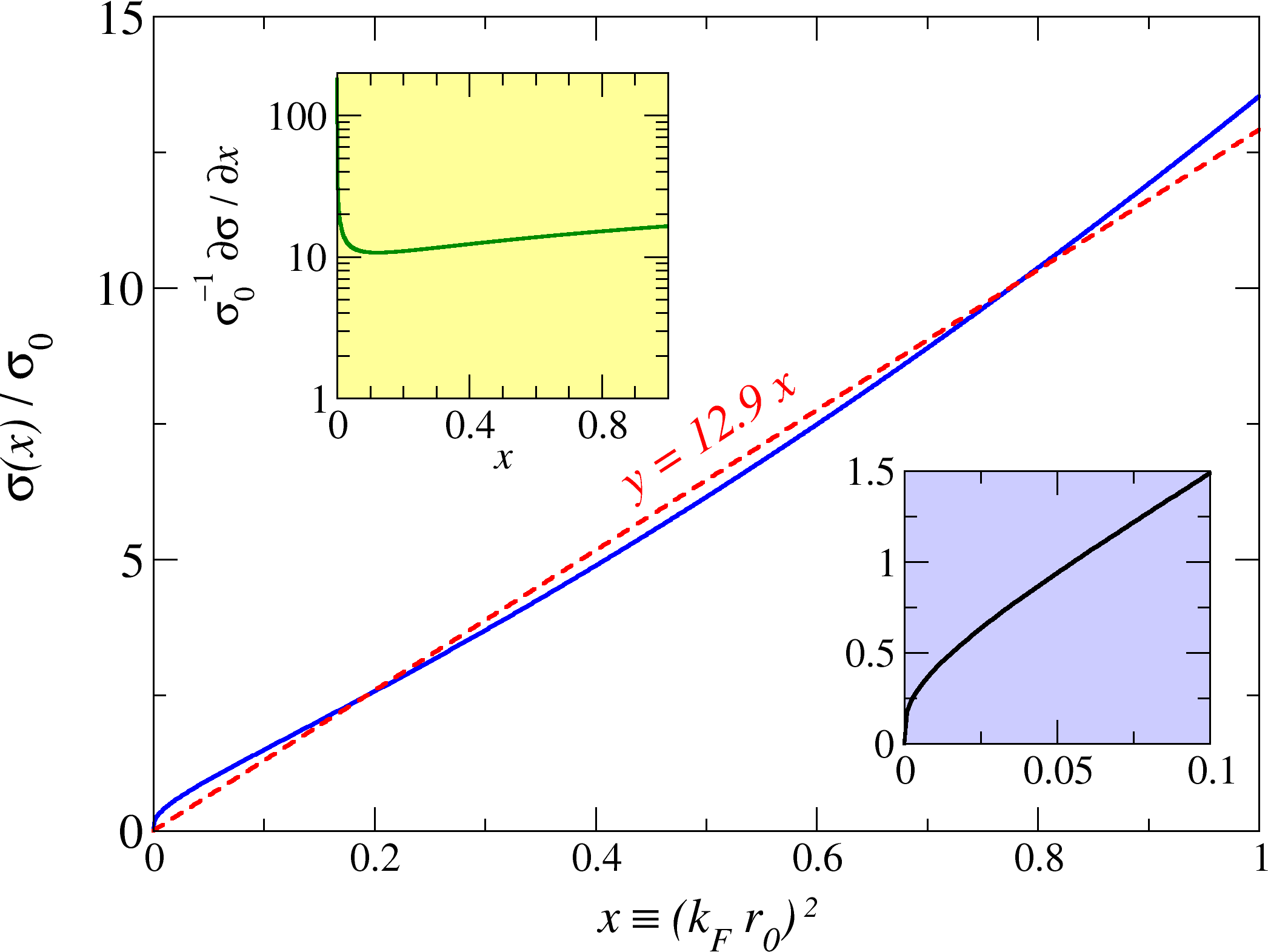}
  \caption{
    \sffamily
    \textbf{DC conductivity under the influence of CS's.}
    DC conductivity \eqref{eq:sigma-r2} versus the adimensional electron 
    density $x\equiv (k_F r_0)^2$. The dashed line shows the best
    linear fit in the entire domain ($y\simeq 12.9 x$).
    Top inset shows $\sigma_0^{-1}\partial\sigma/\partial x$, 
    which is how the electronic mobility ($\mu$) is frequently
    extracted experimentally.
    Bottom inset amplifies the region $k_F\sim0$, dominated
    by a log singularity.
  }
  \label{fig:Meijer}
\end{figure*}

The $1/r^2$ decay in the potential leads to an infrared divergence in $\Phi_\bq$
with a leading order isotropic contribution $\Phi_\bq \approx - 10 \alpha \ce^2
\log(qr_0)$,  all anisotropy being hidden in the subleading terms, with the
regularization distance, $r_0$ of the order of the lattice spacing, reflecting
the relaxation in the neighbourhood of the apex. Then, the CSs scatter primarily
as an isotropic $1/r^2$ potential, and the scattering time for the potential
$\nu_0/r^2$ can be calculated exactly, and reads
\begin{equation}
  \frac{1}{\tau(k_F)} = \frac{2\pi^2 n_i \nu_0^2}{\vF \hbar^2}\,  
    k_F\, G_{2,4}^{3,1}
    \left(4k_F^2r_0^2\Bigl|
      \begin{array}{c}
          -\frac{1}{2},\frac{1}{2}\\0,0,0,-2
      \end{array}\right)
  \label{eq:tau-r2}
  ,
\end{equation}
where $G$ is a Meijer function \cite{Gradshteyn:1965}, $n_i$ the density of
scatterers, and $k_F$ relates to the carrier density via $k_F^2 =\pi n_e$. Then
the longitudinal conductivity follows from \Eqref{eq:tau-r2} and yields
\begin{equation}
  \sigma = \sigma_0 \,
    G_{2,4}^{3,1}\left(4k_F^2r_0^2|
      \begin{array}{c}
        -\frac{1}{2},\frac{1}{2} \\ 0,0,0,-2
     \end{array}\right)^{-1}
  \hspace*{-1em},\quad
  \sigma_0 = \frac{\vF^2\hbar e^2}{2\pi^3n_iv_0^2}\,
  \label{eq:sigma-r2}
\end{equation}
which is only relevant in the regime $0<kr_0\lesssim 1$ shown in
\Fref{fig:Meijer}. We see that  the conductivity is essentially \emph{linear in
electron density} throughout most of the region of interest, except for  the logarithmic singularity around the Dirac point, where it grows sub-linearly. The corresponding
approximate mobility is  $\mu\approx\frac{6v_{F}^{2}\hbar er_0^2}{\pi^2
v_0^2}\frac{1}{n_i}$, and when $ \nu_0$ is replaced by the corresponding
parameter for CSs ($\nu_0\to67\alpha\ce^2/\pi\text{ eV\AA}^2$) one obtains the
mobility for a sea of uncorrelated CSs as $\mu\approx 10^{29} r_0^2 / (n_i
\ce^4) \text{ cm}^2/(\text{Vs})$. Substituting the parameter values $r_0\sim
5\text{ \AA}$ and $n_i \sim10^{12}\text{cm}^{-2}$, results in $\mu\sim10^3/\ce^4
\text{ cm}^2/(\text{Vs})$. The $\ce^4$ dependence reflects a strong sensitivity
to the aperture of the enveloping cone of each CS, but given that $\ce \lesssim
0.5$, it causes relatively small scattering. This effect should thus be more
important in high- mobility suspended samples, where the CSs can become a
limiting factor in carrier mobility.

%
%
\section{Electronic Spectrum}
\begin{figure*}[h]
  \centering
  \includegraphics[width=0.8\textwidth]{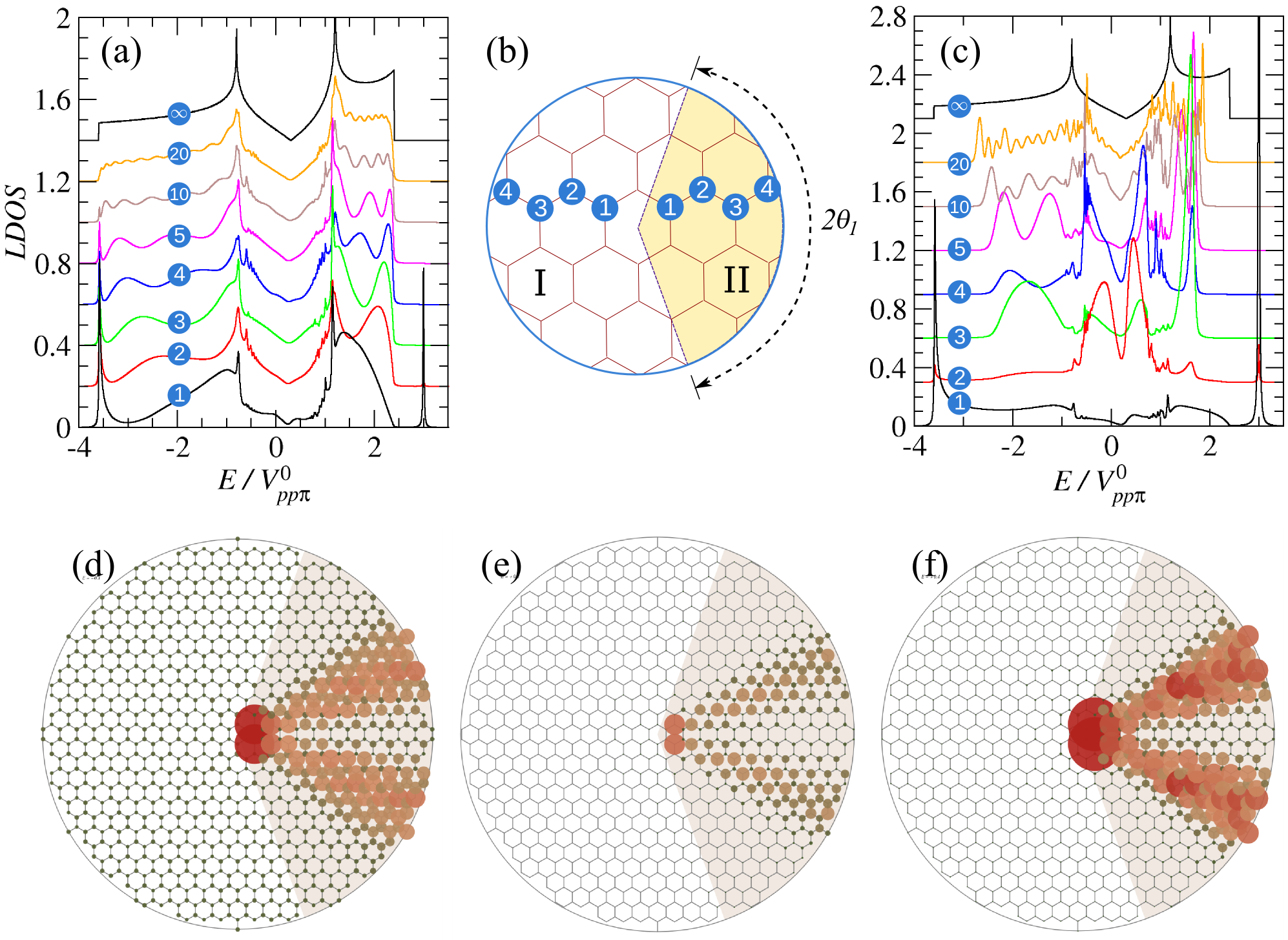}
  \caption{
    \sffamily
    \textbf{LDOS around the apex of a CS.}
    LDOS close to the apex of a CS with $\ce=0.3$, in the regions I,
    \textbf{a}, and II, \textbf{c}, and at specific lattice positions, as
    specified in panel \textbf{b}. 
    The curves in \textbf{a},\textbf{c} were vertically shifted for
    clarity.
    In \textbf{d-f} we show the LDOS in real space for selected energies:
    $E/V_{pp\pi}^0=-0.1,0.3,0.4$. More detailed plots are available 
    in \cite{SuppInfo}.
  }
  \label{fig:LDOS}
\end{figure*}

Although the gauge fields $\Acalb$ are not expected to contribute to transport
at leading order, they do influence the electronic spectrum. In fact, since they
arise from perturbations to nearest neighbour hopping, they might cause
considerable fictitious magnetic fields \cite{Guinea:2009}. To address this at
the level of the lattice, we have calculated the electronic structure associated
with the full tight-binding Hamiltonian \eqref{eq:H-tb} in the presence of a
single unrelaxed CS.
The local density of states [LDOS, $N_\br(E)$] for representative parameters is
shown in \Fref{fig:LDOS}. We see that CSs scatter strongly enough to create
bound electronic states as shown in \Fref{fig:LDOS}(a,c) by the sharp peaks for
states beyond the band edge, decaying rapidly away from the apex. In addition,
the LDOS is very structured at other energies within the band, signalling the
formation of resonant states.This is more clearly visible in \Fref{fig:LDOS}(c)
where the sampling points lie in the region of higher curvature. In this case
the LDOS curves show even stronger perturbation around the Dirac point. The
local bandwidth is decreased, and the leading slope of $N_\br(E)$ around
$E\approx 0$ fluctuates, indicating renormalized Fermi velocities in the
neighbourhood of the apex. In panels \Fref{fig:LDOS}(d-e) we plot the real-space
distribution of the LDOS at representative energies around the Dirac point,
showing that the charge density is mostly localized in the apex, albeit with a
``leak'' along two rays that are at an angle $\approx 24^\text{o}$ with the axis
of symmetry of the CSs, coinciding with the two zero curvature generators  in
the entire conical surface \eqref{eq:psi}, and show clearly the role of
curvature-induced confinement \cite{SuppInfo}. We also always see signals of
``magnetic'' oscillations around the Dirac point, as shown in \Fref{fig:LDOS}
(a)\cite{Endnote-3}. Even though these studies are carried out with zero
magnetic field, the presence of these locally varying fictitious fields is
expected to influence Landau level quantization under a real magnetic field.

%
%
\section{Discussion}
\begin{figure*}[h]
  \centering
  \includegraphics[width=0.8\textwidth]{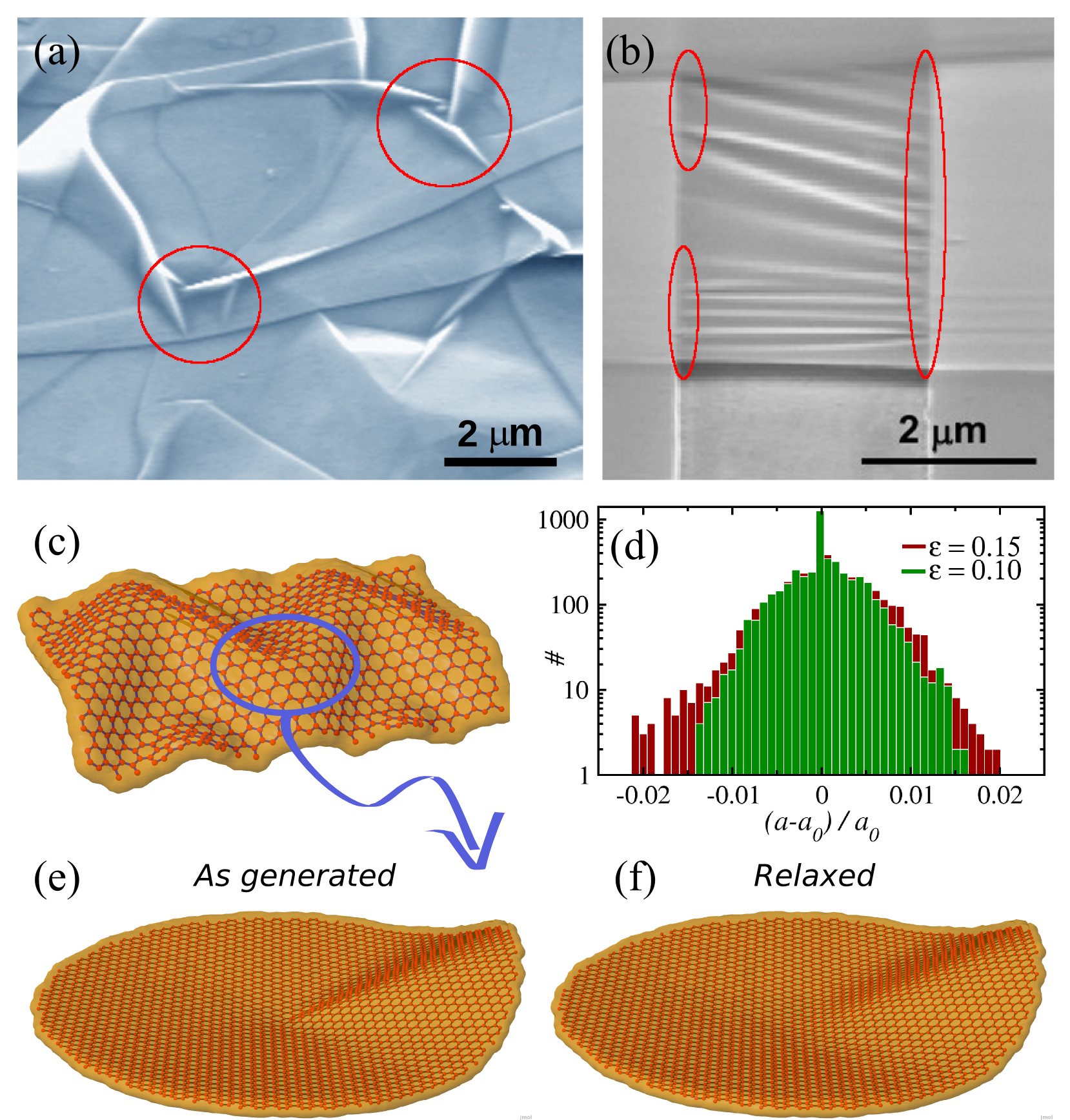}
  \caption{
    \sffamily
    \textbf{Wrinkles in graphene and the origin of conical singularities.}
    \textbf{a}, Folded graphene sheet resembling the draping of a textile, and
    graphene suspended over a trench, \textbf{b} \cite{Endnote-1}.
    Some regions with visible conical singularities (CSs) are highlighted.
    \textbf{c}, Relaxed atomistic profile of a portion of graphene under biaxial
    shear, displaying typical buckling ridges.
    \textbf{d}, Logarithmic histogram of the inter-atomic distances in the
    relaxed configuration \textbf{f} for two values of $\ce$.
    \textbf{e,f}, Profile of the CSs studied here ($\ce=0.1$). The atomic
    positions are shown as generated by applying $\bu(\rho,\theta)$ to all
    atoms, \textbf{e}, and after relaxation by MD, \textbf{f}.
  }
  \label{fig:Fig-1}
\end{figure*}

We have shown that CSs have potential to markedly affect electronic properties
and transport in wrinkled graphene. They contribute a quasi linear-in-density
conductivity, and might even limit mobility in suspended samples at low
temperatures, even in the dilute limit. We did not consider the anisotropy in
the transport calculations, but it
is likely to play a role in situations like \Fref{fig:Fig-1}(b), where  a strong
alignment might lead to coherent scattering. In suspended samples,
curvature-induced disordered flux might be dominant and thus explain why the
quantum Hall effect in 4-terminal suspended samples is so elusive. Our
calculations suggest that it is possible to engineer the electronic and
transport properties in graphene by inducing and controlling CSs on demand using
substrate shear, or by exploring the anomalous thermal expansion coefficient of
graphene, as initiated in \Ref\cite{Lau:2009}. 
Finally, such strong impact of singular deformations on the electronic system
can pave new avenues of interplay between structure and electronics. Graphene,
as seen, has clear and unprecedented advantages, insofar as both its mechanical
response and electronic structure are easily and accurately modelled down to the
atomic level.

%
%
\acknowledgments
AHCN acknowledges the partial support of the U.S. DOE under grant
DE-FG02-08ER46512, and ONR grant MURI N00014-09-1-1063. HL and LM acknowledge
the support of the Harvard-NSF MRSEC and LM acknowledges support from the
MacArthur Foundation.

%
%
\bibliographystyle{apsrev}
\bibliography{graphene_conical}

\begin{thebibliography}{35}
\bibitem[{End({\natexlab{b}})}]{Endnote-3}
\bibinfo{note}{The unrelaxed surface has no in-plain strain, but the 3D
  interatomic distances in the curved lattice do change.}
\end{thebibliography}

%
%

\clearpage
\part*{\Large\centering Supplementary Material}
\bigskip

\section*{Parametrization of conical singularities}

\begin{figure*}[t]
  \centering
  \includegraphics[width=0.8\textwidth]{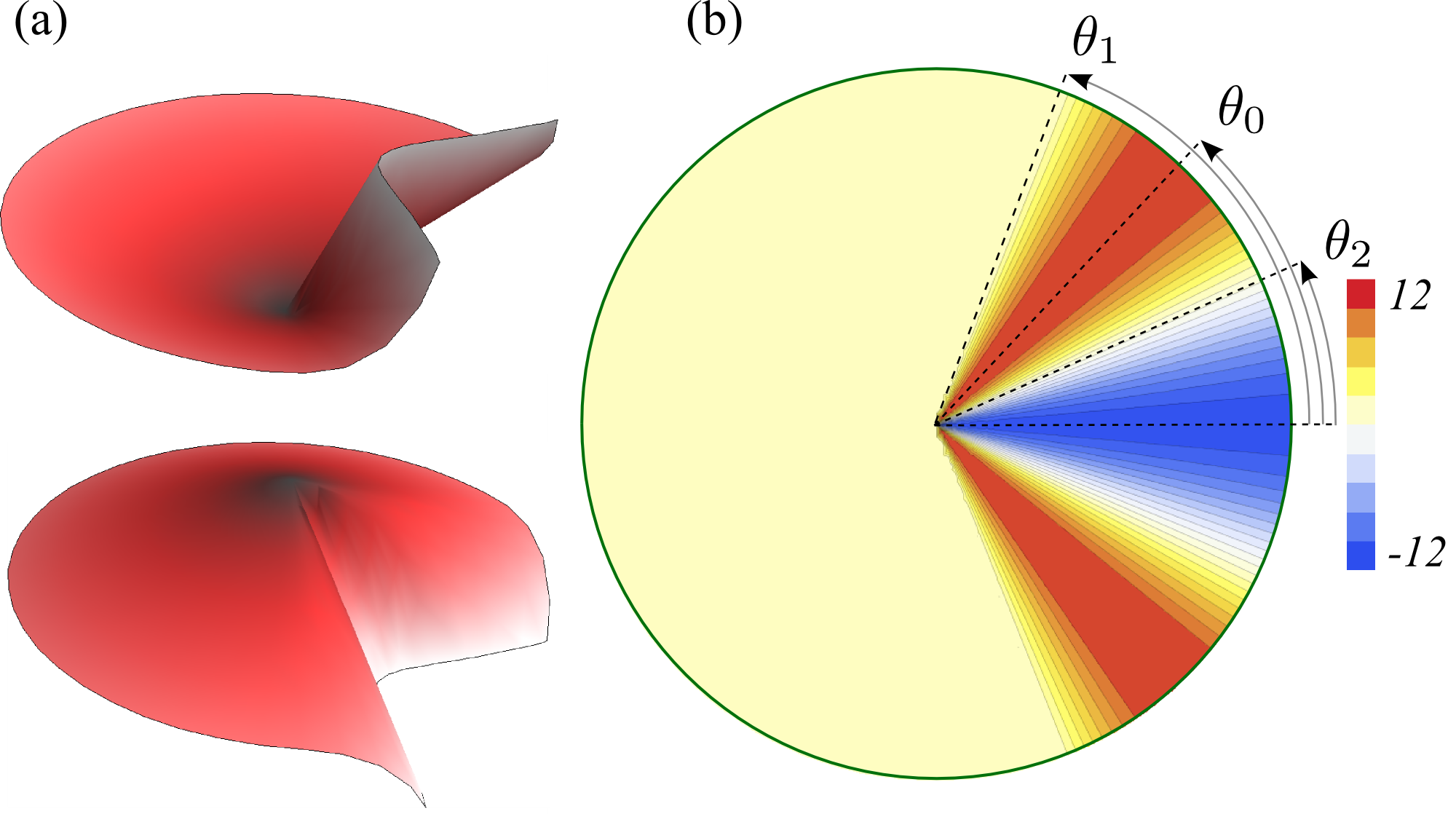}
  \caption{
    \sffamily
    \textbf{Geometry of a conical singularity.}
    \textbf{a}, profile of a conical singularity with $\ce=0.2$ seen   from
    two different perspectives.
    \textbf{b},
    Contour plot of $\ce[\psi''(\theta)+\psi(\theta)]$ revealing
    the curvature as a function of $\theta$. The angles 
    $\theta_1\approx 70^\circ$, $\theta_0\approx 46^\circ$,
    and $\theta_2\approx 24^\circ$ denote the generators along
    which the surface detaches from the enveloping cone, along
    which the curvature is maximal, and along which the curvature is
    null, respectively.
  }
  \label{fig:S4}
\end{figure*}

The fundamental object associated with a crumpled or wrinkled graphene sheet is
the conical singularity \cite{Cerda:1998},  for which there is an analytic
solution that minimizes the elastic energy subject to the constraint of
inextensional deformations. In cylindrical polar coordinates $(\rho, \theta)$,
the displacement field relative to the flat state is given by $\bu(\rho,\theta) 
= \br-\br_0 =u_\rho(\rho,\theta) \bu_\rho + u_\theta(\rho,\theta) \bu_\theta +
\zeta(\rho,\theta) \bz$, where the vertical displacement $\zeta(\rho,\theta) =
\rho \psi(\theta)$ is a generalized cone. The azimuthal component of the
displacement field $\psi(\theta)$ is given by
\begin{equation}
  \psi(\theta) = \ce \Theta(|\theta| - \theta_1) + \ce \Psi(\theta)
\Theta(\theta_1 - |\theta|)
  \label{eq:supp-psi} 
  \,.
\end{equation}
where $ \Theta(\cdot)$ is the usual Heaviside function, the aperture angle of
the enveloping cone is $\pi - 2 \tan^{-1}\ce$, and the universal geometric
function $\Psi(\theta)$ is given by:
\begin{equation}
  \Psi(\theta) = \frac{
      \sin\theta_1 \cos a\theta - a\sin a\theta_1\cos\theta
    }{
      \sin\theta_1 \cos a\theta_1 - a\sin a\theta_1\cos\theta_1
    }
  ,
\end{equation}
where $a\approx 3.8$ and $\theta_1\approx 70^\circ$ are universal numbers. The
general shape of the deformed surface is shown in \Fref{fig:S4}(a). The domain
$|\theta|<\theta_1$ corresponds to the region where the surface coincides with
the cylindrical envelope cone. 

The local azimuthal curvature of the cone is 
$[\psi''(\theta)+\psi(\theta)]/\rho$. In \Fref{fig:S4}(b) we present a contour
plot of $\ce[\psi''(\theta)+ \psi(\theta)]$. For the region $|\theta|>\theta_1$,
the shape is not that of a perfect cone; instead the local mean curvature
increases attaining a maximum at $\theta_0\approx47^\circ$, before decreasing
with an inflection point at $\theta_2\approx24^\circ$ and becomes negative for
$|\theta|<\theta_2$, with a maximum at $\theta =0$.

\section*{Details of the atomistic calculations}

We start with a graphene disk of radius $\sim5nm$ which is geometrically
transformed into a conical shape following the inextensional analytic model of
\cite{Cerda:1998}.  This is done by fixing the coordinates of the atoms of 2
rows near the free edge, while the remaining carbon atoms of the cone are
allowed to relax via a classical molecular dynamics simulation at zero
temperature, so that the analytic solution which is singular at the tip relaxes
by stretching locally. The C-C interactions are modeled with a second generation
reactive empirical bond order (REBO) potential \cite{Brenner:2002}.

\section*{Details of the LDOS calculations}

We use a $\pi$ tight-binding Hamiltonian (2), including overlaps up to
next-to-nearest neighbors on a real Honeycomb lattice of dimensions
$2000\times1000\, a_0^2$. The lattice was deformed to conform to the exact
analytical profile of the conical indentation discussed in the main text. The
apex of the cone lies at the center of the lattice. The hopping integrals
$t_{ij}$ and $t_{ij}^\prime$ were determined from the analytical result (3). The
local density of states (LDOS) was calculated by recursivelly solving for the
local Green's function, until convergence is achieved. The results for the LDOS,
$N_\br(E)$, are hence numerically exact, up to a broadening of $\Delta
E/V_{pp\pi}^0=0.01$, employed for faster convergence.

\section*{Real space distribution of LDOS}

The LDOS, $N_\br(E)$, has been calculated exactly, as discussed above, at all
lattice sites in the vicinity of the apex. for readability, the real-space
images show only a portion -- less than 1\% in area--- of the total lattice.
Since our interest rest on the local electronic structure near the apex, we
considered only one conical singularity. The large lattice size ensures that
spurious edge effects do not interfere with the structure around the apex.

\begin{figure*}[t]
  \centering
  \includegraphics[width=0.8\textwidth]{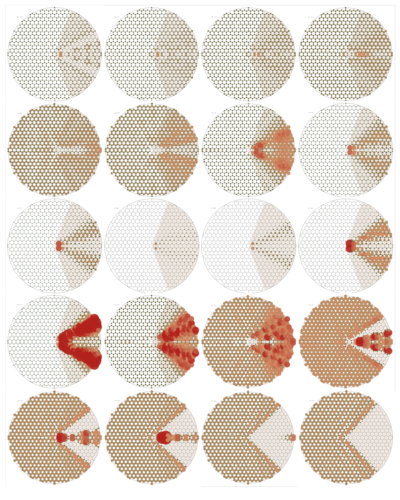}
  \caption{
    \sffamily
    \textbf{LDOS around the apex af a conical singularity.}
    Each panel shows the distribution of the LDOS in real space 
    around the apex of a cone with $\varepsilon=0.3$. Panels pertain
    to energies $E/V_{pp\pi}^0=-1.7,-1.5,-1.3,\dots,1.9,2.1$.
  }
  \label{fig:S1}
\end{figure*}

In \Fref{fig:S1} we present $N_\br(E)$ at all those lattice positions, $\br$,
for a constant energy, $E$. The energy is measured in units of $V_{pp\pi}^0$. At
each lattice site a disk, whose diameter is proportional to the magnitude of the
LDOS, is drawn. The panels in \Fref{fig:S1} pertain to an interval of energies
between $E/V_{pp\pi}^0=-1.7$ and $E/V_{pp\pi}^0=2.1$, at increments of $0.2$.
This interval encompasses the two van Hove singularities, and the Dirac point
which, for our choice of parameters, lies at $E/V_{pp\pi}^0=0.3$.

\begin{figure*}[t]
  \centering
  \includegraphics[width=0.8\textwidth]{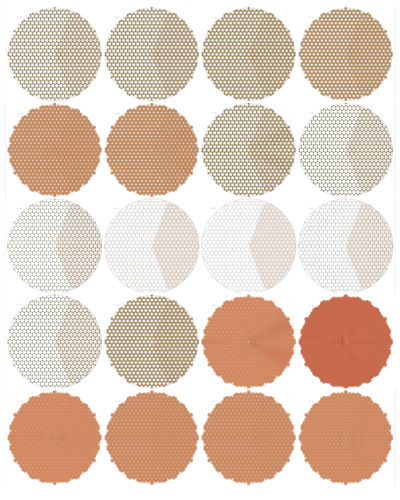}
  \caption{
    \sffamily
    \textbf{LDOS around the apex af a conical singularity.}
    Each panel shows the distribution of the LDOS in real space 
    around the apex of a cone with $\varepsilon=0.1$. Panels pertain
    to energies $E/V_{pp\pi}^0=-1.7,-1.5,-1.3,\dots,1.9,2.1$.
  }
  \label{fig:S2}
\end{figure*}

\begin{figure*}[t]
  \centering
  \includegraphics[width=0.8\textwidth]{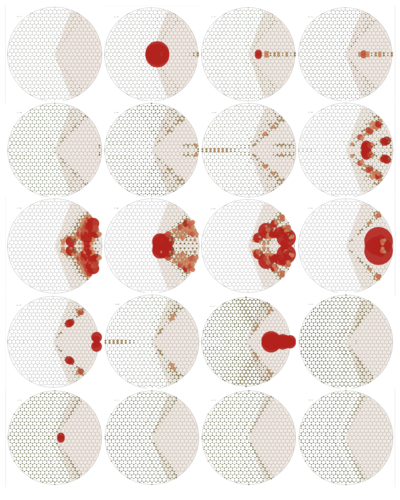}
  \caption{
    \sffamily
    \textbf{LDOS around the apex af a conical singularity.}
    Each panel shows the distribution of the LDOS in real space 
    around the apex of a cone with $\varepsilon=0.5$. Panels pertain
    to energies $E/V_{pp\pi}^0=-1.7,-1.5,-1.3,\dots,1.9,2.1$.
  }
  \label{fig:S3}
\end{figure*}

For comparison, in \Fref{fig:S2} we present precisely the same analysis for a
wider cone, with $\ce=0.1$. In this case, there are no visible features, with
the LDOS remaining greatly homogeneous, and following a variation with energy
much similar to the unperturbed lattice. \Fref{fig:S3} shows an exaggerated case
with $\ce=0.5$. In this case relaxation effects should be stronger and the
approximation of neglecting $\sigma-\pi$ overlaps less warranted. Data is shown
here for perspective only.

\end{document}